\documentclass[aps, prl, twocolumn, superscriptaddress, amsmath,  tightenlines, longbibliography]{revtex4-1}

\usepackage{dcolumn}
\usepackage{graphicx}
\usepackage{mathrsfs}
\usepackage{subfigure}
\usepackage{booktabs}
\usepackage{amsmath}
\usepackage{physics}
\usepackage{dsfont}
\usepackage{amstext}
\usepackage{amssymb}
\usepackage{amsbsy}
\usepackage{bbm}
\usepackage{amsthm}
\usepackage{graphicx}
\usepackage{color}
\usepackage[colorlinks,citecolor=blue]{hyperref}

\usepackage{url}
\usepackage[colorlinks]{hyperref}
\hypersetup{%
	plainpages=true,
	breaklinks=true,       
	hypertexnames=false,  
	pageanchor=true,
	colorlinks=true,
	linkcolor={blue},
	citecolor={red},
	urlcolor={blue},
	anchorcolor={black}
}

 \makeatletter

\newcommand{\Rmnum}[1]{\expandafter\@slowromancap\romannumeral #1@}
\makeatother

\begin{document}

\title{Dissipation and Interaction-Controlled Non-Hermitian Skin Effects}
\author{Yang Li}
\thanks{These authors contributed equally}
\affiliation{School of Physics and Optoelectronics, South China University of Technology,  Guangzhou 510640, China}
\author{Zhao-Fan Cai}
\thanks{These authors contributed equally}
\affiliation{School of Physics and Optoelectronics, South China University of Technology,  Guangzhou 510640, China}
\author{Tao Liu}
\email[E-mail: ]{liutao0716@scut.edu.cn}
\affiliation{School of Physics and Optoelectronics, South China University of Technology,  Guangzhou 510640, China}
\author{Franco Nori}
\affiliation{Center for Quantum Computing, RIKEN, Wako-shi, Saitama 351-0198, Japan}
\affiliation{Department of Physics, University of Michigan, Ann Arbor, Michigan 48109-1040, USA}

\date{{\small \today}}


\begin{abstract}
Non-Hermitian skin effects (NHSEs) have recently been extensively studied at the single-particle level. When many-body interactions become dominant, novel non-Hermitian phenomena can emerge. In this work, we propose an experimentally accessible mechanism to induce and control NHSEs in interacting and reciprocal dissipative systems.  We consider both 1D and 2D Bose-Hubbard lattices subject to staggered two-particle loss combined with synthetic magnetic flux and long-range hopping. When the two-particle loss is small, the bound eigenstates (e.g., doublons and triplons) are all localized at the same boundary due to the interplay between the magnetic flux and staggered two-particle loss. In contrast, for strong two-particle loss, the skin-mode localization direction of the bound particles is unexpectedly reversed. This reversal stems from the combined effect of the staggered two-particle loss, synthetic magnetic flux, and long-range hopping, through which virtual second-order and third-order hopping processes induce effectively strong nonreciprocal hopping of doublons. Our results open up a new avenue for exploring novel non-Hermitian phenomena in many-body systems. 
\end{abstract}

\maketitle

\textit{{\color{blue}Introduction}}.---In the framework of non-Hermitian quantum mechanics \cite{Ashida2020}, open quantum systems described by effective non-Hermitian Hamiltonians exhibit many striking physical phenomenon without Hermitian counterparts \cite{Peng2014a, Peng2014b, Gao2015, PhysRevLett.118.040401, PhysRevLett.118.045701,arXiv:1802.07964,El-Ganainy2018,ShunyuYao2018,PhysRevLett.125.126402,PhysRevLett.123.066404, YaoarXiv:1804.04672,PhysRevLett.121.026808,PhysRevLett.122.076801,PhysRevLett.123.170401, PhysRevLett.123.206404,PhysRevLett.123.066405,PhysRevLett.123.206404, ZhangJ2018,Bliokh2019, PhysRevB.100.054105, PhysRevA.100.062131,Zhao2019,Bliokh2019,PhysRevX.9.041015,PhysRevLett.124.056802, PhysRevB.97.121401,    PhysRevA.101.062112,PhysRevA.102.033715,  PhysRevLett.124.086801, PhysRevLett.127.196801, RevModPhys.93.015005,PhysRevLett.128.223903,  Zhang2022, PhysRevLett.129.093001,arXiv:2505.05058, Ren2022,PhysRevX.13.021007,PhysRevLett.131.036402,PhysRevLett.131.116601,Leefmans2022,haowang2025, Okuma2023,arXiv:2403.07459,arXiv:2311.03777,PhysRevLett.132.070402,Manna2023,PhysRevA.109.063329,Parto2023, PhysRevX.14.021011,PhysRevLett.132.050402,arXiv:2411.10398,Leefmans2024,PhysRevLett.133.136602,Guo2024,Roccati2024}. One of the fascinating properties of non-Hermitian systems is the localization of bulk modes at the boundaries \cite{ShunyuYao2018,PhysRevLett.125.126402,PhysRevLett.123.066404, YaoarXiv:1804.04672,PhysRevLett.121.026808,PhysRevLett.122.076801,PhysRevLett.123.170401, PhysRevLett.123.206404,PhysRevLett.123.066405,PhysRevLett.123.206404}, dubbed the non-Hermitian skin effect (NHSE). The NHSE originates from the intrinsic  point-gap topology of complex eigenenergies \cite{ PhysRevLett.124.086801}. Recently, many efforts have been devoted to exploring unique consequences of NHSE  without many-body interactions, e.g., breakdown of conventional Bloch band theory \cite{ShunyuYao2018,PhysRevLett.123.066404,PhysRevLett.125.126402}    and entanglement phase transitions \cite{PhysRevX.13.021007}. 

When many-body interactions are introduced into non-Hermitian systems, unusual physical properties can emerge \cite{Ashida2017, PhysRevLett.121.203001,PhysRevLett.123.123601,PhysRevLett.124.147203,PhysRevB.104.075106,PhysRevB.106.L121102,PhysRevB.106.235125,PhysRevB.105.165137, PhysRevB.102.235151,PhysRevB.106.205147,Shen2022,PhysRevLett.129.180401,PhysRevResearch.5.033173,Longhi2023,PhysRevA.107.043315,PhysRevLett.123.090603,PhysRevB.109.L140201, PhysRevLett.132.096501,PhysRevLett.133.076502,arXiv:2405.1228,arXiv:2403.10449,Kim2024,PhysRevLett.132.120401,arXiv:2408.12451,PhysRevB.110.045440,PhysRevB.111.205418,PhysRevB.109.155434,PhysRevLett.133.216601,PhysRevLett.133.136503,PhysRevLett.132.116503,Shen2025}. Especially, with nonreciprocal hopping, many-body interactions have the potential to induce  NHSE   \cite{PhysRevLett.129.180401},   occupation-dependent NHSE \cite{PhysRevLett.132.096501},   non-Hermitian Mott skin effect \cite{PhysRevLett.133.076502},  dynamical suppression of NHSE \cite{arXiv:2405.1228},  and so on. However,  it is challenging to implement the nonreciprocal hopping in order to achieve  the NHSE for many experimental platforms.   A more operational approach is to utilize onsite loss   \cite{PhysRevLett.124.250402, PhysRevLett.125.186802, PhysRevLett.128.223903, PhysRevLett.129.070401,PhysRevA.109.063329}. Therefore, a natural open question arises: can the onsite loss   be utilized to induce and even control NHSEs in the interacting non-Hermitian system?

\begin{figure}[!b]
	\centering
	\includegraphics[width=8.7cm]{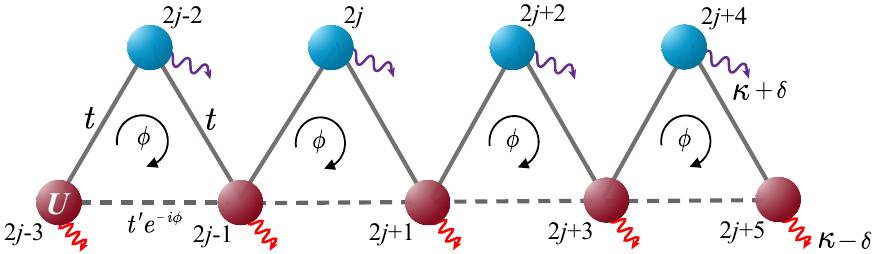}
	\caption{Schematic showing a 1D Bose-Hubbard model with long-range hopping, arranged in a two-leg triangular configuration. The onsite interaction is denoted by $U$, while  staggered two-particle losses between odd and even sites, with rates $\kappa \pm \delta$, are indicated by red and purple wavy lines. The parameters $t$ and $t^{\prime}$ represent the nearest-neighbor (dark solid lines) and next-nearest-neighbor (dark dashed lines) hopping amplitudes, respectively. The complex next-nearest-neighbor hopping introduces a synthetic magnetic flux $\phi$ that threads each triangular plaquette. }\label{Fig1}
\end{figure}

In this work, we propose to utilize two-particle dissipation to induce and control the NHSE in interacting and reciprocal systems. We construct and investigate both one-dimensional (1D) and two-dimensional (2D) Bose-Hubbard models incorporating staggered two-particle loss, synthetic magnetic flux, and next-nearest-neighbor long-range hopping. For small two-particle loss, all the bound particles are localized at the boundaries, indicating the occurrence of the NHSE. In contrast to the case with staggered single-particle loss or multi-path interference studied previously, strong two-particle loss unexpectedly reverses the localization direction of the skin modes. We reveal that this hidden mechanism stems from the interplay of staggered two-particle loss, synthetic magnetic flux, and long-range hopping of doublons, which generates effectively strong nonreciprocal hopping via virtual second- and third-order processes.

\begin{figure*}[!tb]
	\centering
	\includegraphics[width=17.8cm]{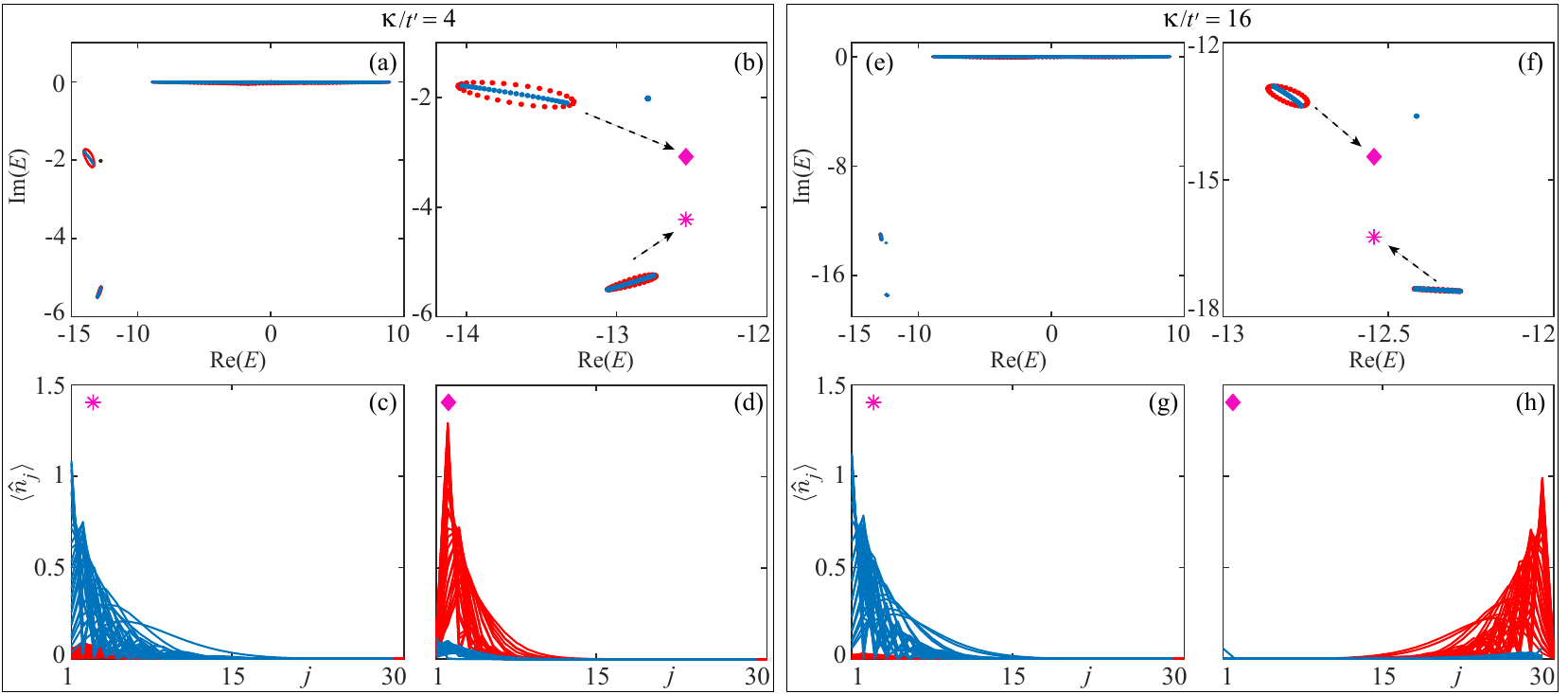}
	\caption{ Complex eigenenergies of $\hat{\mathcal{H}}_\text{nH}$, in triangular configuration, under OBCs (blue dots) and PBCs (red dots) (a) for $\kappa/t'=4$ and  (e) for $\kappa/t'=16$.  The zoom‐in view of   two doublon bands on the far left of (a,e) is shown in (b,f), where the magenta star and diamond denote the two branches of the doublon bands. The corresponding particle densities $\langle \hat{n}_j \rangle$ are shown in 	(c,d) and (g,h) under OBC, where the  red (blue) curves indicate the state  distributions at the odd (even) sites.  Other parameters are  $ U/t' = 6 $,  $ t/t'=2 $,   $\delta/t'=2$, $\phi = \pi/2$, and $L=59$.}\label{FigTrig2}
\end{figure*}

\textit{{\color{blue}Model}}.---Either staggered single-particle loss in reciprocal systems or interactions coupled with a dynamical gauge field can induce the NHSE \cite{PhysRevLett.124.250402, PhysRevLett.125.186802, PhysRevLett.128.223903, PhysRevA.109.063329, PhysRevLett.129.180401}. Interestingly, interference between multiple hopping pathways in a nonreciprocal ladder can even reverse the NHSE \cite{PhysRevB.106.085427, PhysRevLett.132.096501}. Here, we propose a distinct mechanism to induce and control the NHSE: staggered two-particle loss combined with synthetic magnetic flux and long-range hopping in interacting and reciprocal systems. This experimentally accessible configuration enables the realization of effectively strong nonreciprocal hopping in the strong dissipation regime in spite of reciprocal interacting systems, providing a fundamentally different route from previous approaches.

To implement this approach, we first consider a one-dimensional (1D) Bose-Hubbard model with long-range hopping. As shown in Fig.~\ref{Fig1}, the 1D lattice is arranged in a triangular configuration, featuring   next-nearest-neighbor hopping, onsite many-body interactions, and two-particle dissipation. The phase associated with the long-range hopping introduces a synthetic magnetic flux $\phi$, which threads through each triangular plaquette. When including the staggered two-particle loss $\kappa \pm \delta$ on odd and even sites, the effective non-Hermitian Hamiltonian of the system  is written as
\begin{align}\label{HermitianH}
	\hat{\mathcal{H}}_\text{nH} = & -\sum_{j} \left(t \hat{a}^\dagger_{j+1} \hat{a}_j + t^\prime e^{-i\phi} \hat{a}^\dagger_{2j+1} \hat{a}_{2j-1} + \text{H.c.}\right) \nonumber \\ &   - U  \sum_{j=1}^L \hat{a}^\dagger_j \hat{a}^\dagger_j \hat{a}_j \hat{a}_j  - \frac{i}{2} \sum_{j=1}^{L} \left[\kappa + (-1)^{j} \delta  \right] \hat{a}^\dagger_j \hat{a}^\dagger_j \hat{a}_j \hat{a}_j,
\end{align}
where $\hat{a}_j$ annihilates a boson at site $j$, $t$ and $t^\prime$ represent  the nearest-neighbor and next-nearest-neighbor hopping strengths, $U $ denotes the onsite many-body interacting strength, and $L$ is the number of lattice sites. The imaginary onsite many-body interaction in the last term can be realized by
continuously monitoring the particle number followed by a postselection measurement (see details in Sec.~\Rmnum{1} of Supplementary Material (SM) in Ref.~\cite{NonHermitianDoublon2024}).

Since $[\hat{\mathcal{H}}_\text{nH}, ~\sum_{j} \hat{n}_j] = 0$, with $\hat{n}_j = \hat{a}_j^\dagger \hat{a}_j$, the particle number is conserved, allowing analysis within fixed-number subspaces.

\textit{{\color{blue}Manipulating NHSE in Double-Excitation Subspace}}.---We begin by investigating the control of the NHSE of doublons within two-particle excitation subspace. We plot the complex  eigenenergies of $\hat{\mathcal{H}}_\text{nH}$ for different      $\kappa$ under open boundary conditions (OBCs, blue dots) and periodic boundary conditions (PBCs, red dots) in Fig.~\ref{FigTrig2}(a,e). 
	
The eigenspectrum consists of a continuum of scattering states and discrete doublon bands. The scattering states are superpositions of two particles on different sites, while the doublon bands correspond to bound pairs occupying the same site, with eigenenergies well separated from the scattering continuum [see Fig.~\ref{FigTrig2}(a,e)]. We focus on two such doublon bands, highlighted on the far left of Fig.~\ref{FigTrig2}(a,e) and magnified in Fig.~\ref{FigTrig2}(b,f). In the complex plane, each doublon band, marked by a magenta star or diamond, forms a point gap (red dots) under PBCs. These point gaps enclose the corresponding doublon modes (blue dots) under OBCs, suggesting the presence of the NHSE. To confirm this, we compute the single-particle occupation density as
\begin{align}\label{density}
	\langle \hat{n}_j \rangle = \frac{_R\!\bra{\psi_m} \hat{n}_j  \ket{\psi_m}_R }{_R\!\bra{\psi_m} \ket{\psi_m}_R}, 
\end{align}
where $\ket{\psi_m}_R$ ($m=1,2,3,\cdots$) is the $m$th right eigenvector with $\hat{\mathcal{H}}_\text{nH} \ket{\psi_m}_R = E_m \ket{\psi_m}_R$.

\begin{figure}[!tb]
	\centering
	\includegraphics[width=8.7cm]{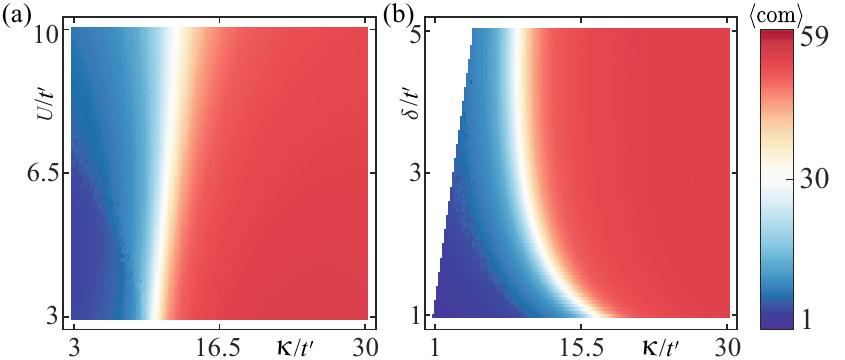}
	\caption{ Phase diagram of reversible skin modes within the doublon band with larger imaginary part in (a)  $(U, \kappa)$ plane for $\delta/t' = 2$ and $t/t' = 2$, and (b)  $(\delta, \kappa)$ plane for $U/t' = 6$, $\phi = \pi/2$, and $t/t' = 2$.  }\label{FigTrig_phaseDiagram}
\end{figure}

Figures \ref{FigTrig2}(c,d) and \ref{FigTrig2}(g,h) show the site-resolved particle densities of the two doublon bands for $\kappa/t' = 4$ and $\kappa/t' = 16$, respectively.  The doublons originating from two distinct bands become localized at the boundary due to the NHSE,  predominantly occupying the odd and even sites, respectively.   Most remarkably, large two-particle dissipation can reverse the skin-mode localization direction of one doublon band [see Fig.~\ref{FigTrig2}(d,h)].   Note that the the NHSE of doublons originates from intrinsic non-Hermitian topology, which is characterized by the winding number (see details in End Matter).

Previous studies have shown that staggered single-particle loss, combined with magnetic flux, can induce the NHSE \cite{ PhysRevLett.124.250402,   PhysRevLett.129.070401}, but it cannot reverse the localization direction even for the interacting non-Hermitian case, as demonstrated in Sec.~\Rmnum{2} of  SM \cite{NonHermitianDoublon2024}.  	These results demonstrate that the staggered two-particle loss, when combined with long-range hopping and magnetic flux,   constitutes a fundamentally novel mechanism for inducing and reversing the NHSE, distinct from previously reported scenarios, such as the multiple-hopping-pathway interference mechanism proposed for NHSE reversal \cite{PhysRevB.106.085427, PhysRevLett.132.096501}.

To determine the parameter regime for reversible skin modes, we calculate the averaged center of mass (com) within a single doublon band, defined as
\begin{align}\label{aveapos}
	\langle	\textrm{com} \rangle =  \Bigg \langle\frac{\sum_j j \langle \hat{n}_j \rangle}{\sum_j \langle \hat{n}_j \rangle} \Bigg \rangle_E, 
\end{align}
where $\langle \cdot \rangle_E$ denotes the average over all eigenstates within a single doublon band. For $\langle \textrm{com} \rangle \sim 1$, it indicates left-edge skin-mode localization, while $\langle \textrm{com} \rangle \sim L$ indicates right-edge skin-mode localization.

Figure \ref{FigTrig_phaseDiagram} shows the phase diagrams of reversible skin modes within the doublon band with the larger imaginary part. The dark blue ($\langle \textrm{com} \rangle \sim 1$) and red ($\langle \textrm{com} \rangle \sim L$) regions indicate skin-mode localization at the left and right edges, respectively, while the  blue ($\langle \textrm{com} \rangle \sim L/2$) regions correspond to critical phases with extended states. These results indicate that the skin-mode localization of doublon can be reversed by tuning the staggered two-particle dissipation $\kappa \pm \delta$ only for a finite $U$ and $\delta$.

\begin{figure}[!b]
	\centering
	\includegraphics[width=8.7cm]{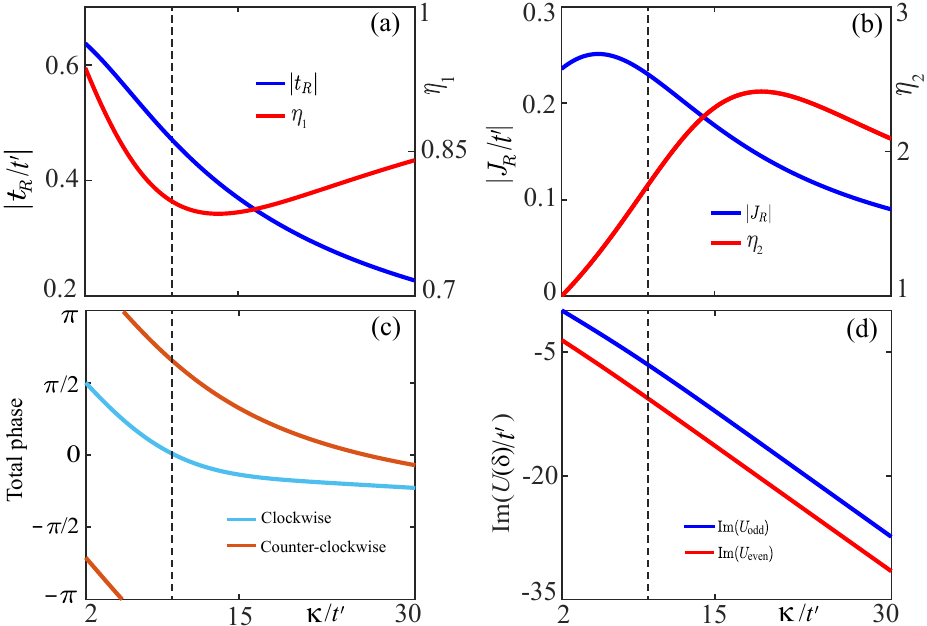}
	\caption{ (a) $\abs{t_R/t'}$ and $\eta_1 = \abs{t_R/t_L}$ versus $\kappa/t'$, (b) $\abs{J_R/t'}$ and $\eta_2 = \abs{J_R/J_L}$ versus $\kappa/t'$, (c)  total phase threading each triangular plaquette versus $\kappa/t'$, and (d) $U(\delta)/t'$   versus $\kappa/t'$. The black dashed vertical line marks the position of the reversed skin mode. Other parameters are  $ U/t' = 6 $,  $ t/t'=2 $, $\phi = \pi/2$, and  $\delta/t'=2$.}\label{FigTrig3}
\end{figure}


 \begin{figure*}[!tb]
	\centering
	\includegraphics[width=17.8cm]{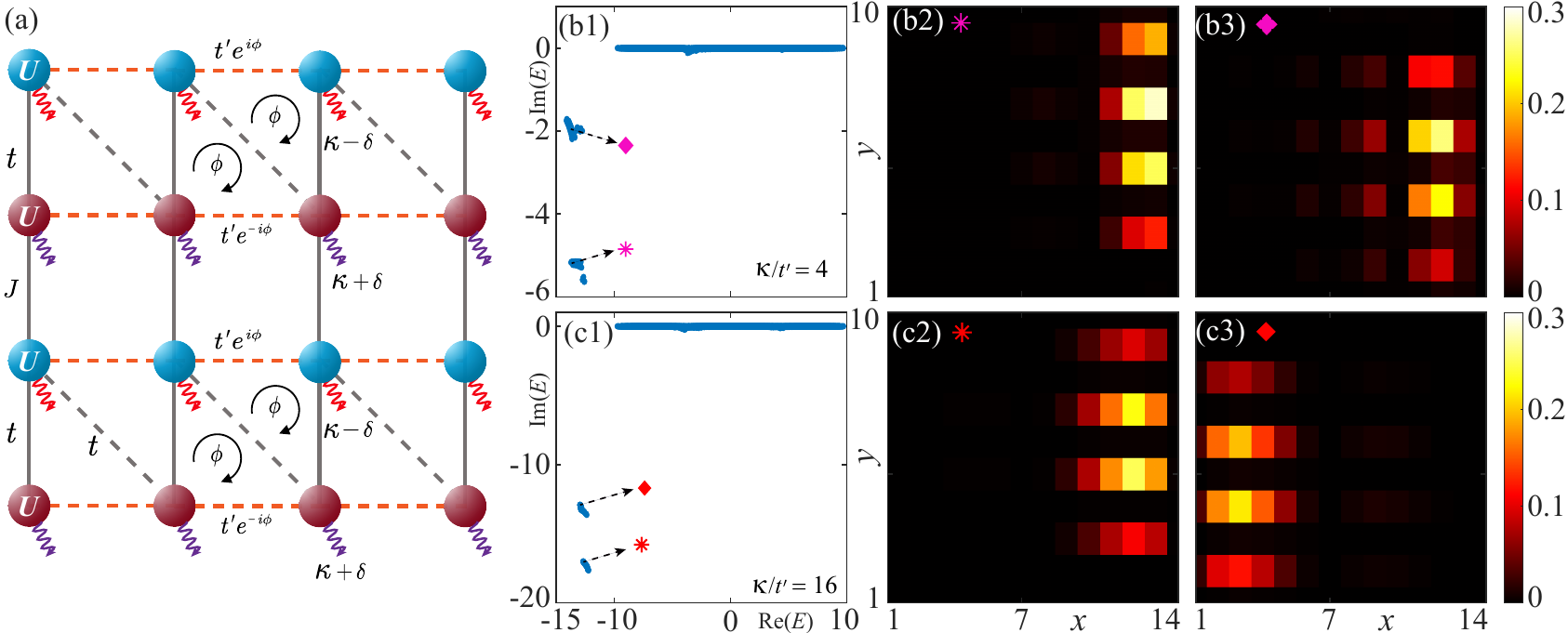}
	\caption{ (a) Schematic showing a 2D Bose-Hubbard model with long-range hopping (dark dashed lines) in a square lattice. The complex hopping (red dashed lines) along the $x$ direction introduces a synthetic magnetic flux $\phi$ that threads each triangular plaquette. The hopping and two-particle loss is staggered along the $y$ direction. Complex eigenenergies of $\hat{\mathcal{H}}_\text{2D}$ and corresponding particle densities $\langle \hat{n}{(x,y)} \rangle$ under OBC: (b1–b3) for $\kappa/t' = 4$ and (c1–c3) for $\kappa/t' = 16$. The magenta star and diamond denote the two branches of the doublon bands. Other parameters are $U/t' = 6$, $t/t' = 2$, $J/t' = 1$, $\phi = \pi/2$, and $\delta/t' = 2$.}\label{Fig2DLattice}
\end{figure*}

\textit{{\color{blue}Effective Hamiltonian of Doublons}}.---To  understand the underlying hidden mechanism of the unexpected reversal of the  localization direction of the doublons, we derive the effective Hamiltonian $\hat{\mathcal{H}}_\text{eff}$ describing the doublon bands in the strong-interaction limit with   $\abs{U} \gg \abs{t}, \abs{t^\prime}$. In this case, the doublons are tightly bound, and  their bands are energetically well-separated from   two-particle scattering states. Therefore, we can split the Hamiltonian $\hat{\mathcal{H}}_\text{nH}$ into its dominant term, with $\hat{\mathcal{H}}_0 = -  \sum_{j} [U+i(\kappa + (-1)^{j} \delta)/2] \hat{n}_j (\hat{n}_j-1)$, and treat the remaining terms as the perturbation. Based on  third-order quasidegenerate  perturbation theory (See details in Sec.~\Rmnum{3} of SM \cite{NonHermitianDoublon2024}), we obtain the effective Hamiltonian $\hat{\mathcal{H}}_\text{eff}$ of doublons as
\begin{align}\label{Hamileff20}
	\hat{\mathcal{H}}_\text{eff} = 	& -  \sum_{j=1}^{L-1} \left( t_R \hat{d}^\dagger_{j+1} \hat{d}_j +  t_L \hat{d}^\dagger_{j} \hat{d}_{j+1}\right) \nonumber \\
	& - \sum_{j=1}^{L/2-2} \left(J_\text{R} \hat{d}^\dagger_{2j+1} \hat{d}_{2j-1} + J_\text{L} \hat{d}^\dagger_{2j-1} \hat{d}_{2j+1} \right)  \nonumber \\
	& + U_1 \sum_{j=1}^{L/2} \hat{d}^\dagger_{2j-1} \hat{d}_{2j-1} + U_2 \sum_{j=1}^{L/2} \hat{d}^\dagger_{2j} \hat{d}_{2j},
\end{align}
where $t_\text{L}=t(\phi)$, $t_\text{R}=t(-\phi)$, $J_\text{L}=J(\phi)$, $J_\text{R}=J(-\phi)$, $U_1 = 	U(\delta)$,   $U_2 = U(-\delta)$, and  
\begin{align}\label{tLBulkeffectnew0}
	t(\phi) = \frac{2 (2U+i\kappa) t^2}{(2U+i\kappa)^2+\delta^2} + \frac{6\left[ (2U+i\kappa)^2-\delta^2 \right] t^2 t^\prime e^{i\phi} }{\left[(2U+i\kappa)^2 + \delta^2 \right]^2},
\end{align}
\begin{align}\label{JRoddeffectspy0}
	J(\phi) =  \frac{6t^2t^\prime }{  (2U+i\kappa- i\delta )^2} e^{i\phi} +  \frac{2 t^{\prime2} }{2U+i\kappa - i\delta} e^{2i\phi},
\end{align}
\begin{align}\label{UoddBulkeffect20}
	U(\delta)   = & -2U -i(\kappa-\delta) - \frac{4(t^2 + {t^\prime}^2)}{2U+i\kappa -i\delta} \nonumber \\ &  - \frac{8 t^2 t^\prime \cos \phi }{(2U+i\kappa - i\delta)^2}.
\end{align}

According to Eq.(\ref{JRoddeffectspy0}), the virtual second-order hopping process of particle pairs contributes an effective phase of $2\phi$, while the third-order process, which requires both nearest-neighbor hopping $t$ and next-nearest-neighbor hopping $t'$, contributes a phase of $\phi$ (see details in Ref.\cite{NonHermitianDoublon2024}). Therefore, \textit{the staggered two-particle loss, combined with synthetic magnetic flux and long-range hopping, can induce strong nonreciprocal hopping of doublons—with $\abs{J_{\text{R}}} \neq \abs{J_{\text{L}}}$—in the interacting and reciprocal system through appropriate design of the system parameters}.

Figure \ref{FigTrig3} shows $\abs{t_R}/t'$ and its ratio $\eta_1 = \abs{t_R / t_L}$, $\abs{J_R}/t'$ and its ratio $\eta_2 = \abs{J_R / J_L}$, the effective magnetic flux threading the triangular lattice of doublons, and the effective staggered onsite loss versus $\kappa/t'$. Most remarkably, the effective next-nearest-neighbor hopping of doublons becomes strongly nonreciprocal as $\kappa$ increases [see Fig.~\ref{FigTrig3}(b)]. For small $\kappa$, the nearest-neighbor and next-nearest-neighbor hoppings are only slightly nonreciprocal. In this regime, the left-edge skin modes are caused primarily by the effective magnetic flux and the staggered onsite loss at even and odd sites. However, for large $\kappa$, the next-nearest-neighbor hopping along the odd sites becomes highly nonreciprocal, with a larger right hopping strength $J_R$, while the nearest-neighbor hopping remains only slightly nonreciprocal. Therefore, the doublon with the larger imaginary part (which mainly occupies the odd sites, as shown in Fig.~\ref{FigTrig2}) becomes reversely localized toward the right edge.

The proposed mechanism can control the NHSE not only in the triangular lattice but also in the zigzag lattice (see details in Sec.~\Rmnum{4} of SM \cite{NonHermitianDoublon2024}). In addition to two-particle excitations, the reversal of the NHSE can also occur in higher-excitation subspaces (see details in End Matter).

\textit{{\color{blue}Control NHSE in 2D}}.---The staggered two-particle loss, combined with synthetic magnetic flux and long-range hopping, can also induce and control the NHSE in a 2D interacting system. As shown in Fig.~\ref{Fig2DLattice}(a), we consider a 2D Bose-Hubbard model with staggered two-particle loss $\kappa \pm \delta$, magnetic flux $\phi$, and long-range hopping (dark dashed lines) on a square lattice. The effective non-Hermitian Hamiltonian is written as
\begin{align}\label{Hermitian2D}
	\hat{\mathcal{H}}_\text{2D}  = & -\sum_{x,y} \left(t \hat{a}^\dagger_{x,y+1} \hat{a}_{x+1,y} + t^\prime e^{i(-1)^y\phi} \hat{a}^\dagger_{x+1,y} \hat{a}_{x,y} + \text{H.c.}\right) \nonumber \\ & 
	-\sum_{x,y} \left(t \hat{a}^\dagger_{x,2y} \hat{a}_{x,2y-1} + J \hat{a}^\dagger_{x,2y+1} \hat{a}_{x,2y} + \text{H.c.}\right) \nonumber \\ &   -   \sum_{x,y}  \left[U + \frac{i}{2}\big(\kappa-(-1)^y \delta\big)\right] \hat{a}^\dagger_{x,y} \hat{a}^\dagger_{x,y} \hat{a}_{x,y} \hat{a}_{x,y} 	,
\end{align}

Figure \ref{Fig2DLattice} shows the complex eigenenergies of $\hat{\mathcal{H}}_\text{2D}$ and the corresponding particle densities $\langle \hat{n}(x,y) \rangle$ under OBC. For small $\kappa$, the doublons, indicated by the magenta star and diamond, are all localized at the right boundary [see Fig.~\ref{Fig2DLattice}(b1)–(b3)]. However, for large $\kappa$, the localization direction of the skin modes within the doublon band with larger imaginary part is reversed [see Fig.~\ref{Fig2DLattice}(c3)] due to the effectively strong nonreciprocal hopping induced by the staggered two-particle loss combined with synthetic magnetic flux and long-range hopping. Therefore, this proposal represents a novel mechanism for inducing and controlling the NHSE in both 1D and 2D reciprocal interacting systems.

\textit{{\color{blue}Conclusion}}.---We construct both 1D and 2D Bose-Hubbard models to demonstrate a novel dissipative mechanism for controlling the NHSE. These interacting and reciprocal systems incorporate staggered two-particle loss, synthetic magnetic flux, and next-nearest-neighbor long-range hopping. We show that, beyond the emergence of the NHSE for bound particles (e.g., doublons and triplons) under weak dissipation, the localization direction reverses as the two-particle loss increases. This control stems from effectively strong nonreciprocal hopping of doublons, generated by the interplay of staggered two-particle loss, synthetic magnetic flux, and long-range hopping via virtual higher-order processes. Such a mechanism is absent in systems with staggered single-particle loss or without interactions. Our work uncovers a novel, experimentally accessible route to induce and control the NHSE.

\begin{acknowledgments}
T.L. acknowledges the support from  National Natural	Science Foundation of China (Grant No.~12274142), Introduced Innovative Team Project of Guangdong Pearl River Talents Program (Grant
No. 2021ZT09Z109), the Fundamental Research Funds for the Central Universities (Grant No.~2023ZYGXZR020),  and the Startup Grant of South China University of Technology (Grant No.~20210012).  F.N. is supported in part by:   the Japan Science and Technology Agency (JST)[via the Quantum Leap Flagship Program (Q-LEAP), and the Moonshot R$\&$D Grant Number JPMJMS2061], the Asian Office of Aerospace Research and Development (AOARD) (via Grant No.~FA2386-20-1-4069), and the Office of Naval Research (ONR) Global (via Grant No.~N62909-23-1-2074).
\end{acknowledgments}

\section{End Matter}

\begin{figure*}[!bt]
	\centering
	\includegraphics[width=17cm]{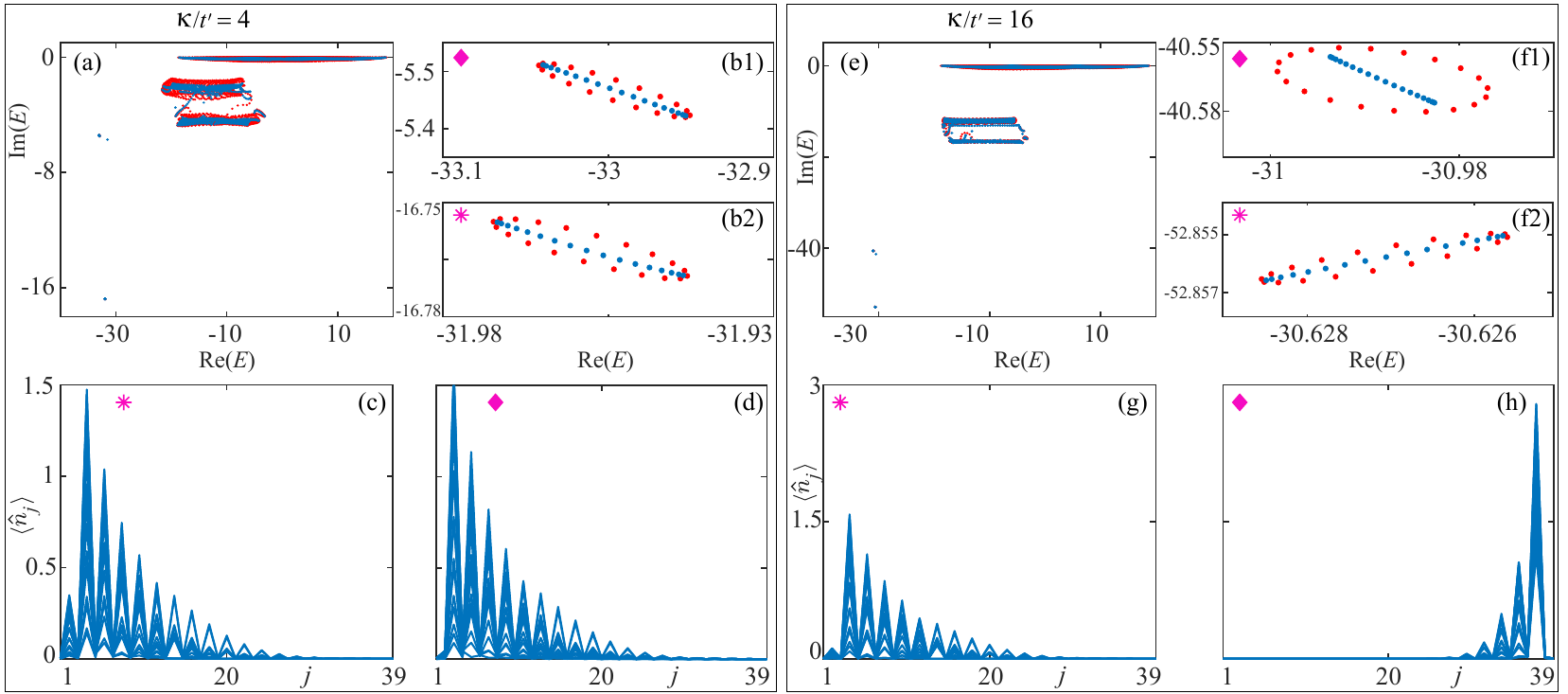}		
	\caption{Complex eigenenergies of $\hat{\mathcal{H}}_\text{nH}$ within three-excitation subspace, in the triangular configuration, under OBCs (blue dots) and PBCs (red dots) (a) for $\kappa/t'=4$ and  (e) for $\kappa/t'=16$.  The zoom‐in view of the two doublon bands on the far left of (a,e) is shown in (b1,b2,f1,f2), where the magenta star and diamond denote the two branches of the doublon bands. The corresponding particle densities $\langle \hat{n}_j \rangle$ are shown in 	(c,d) and (g,h) under OBC.  Other parameters are  $ U/t' = 5 $,  $ t/t'=3 $,  $\delta/t'=2$, $\phi = \pi/2$, and $L=39$.}\label{FigTrigThree}
\end{figure*}

\textit{{\color{blue}Manipulating NHSE in Higher-Excitation Subspace}}.---In the main text, we demonstrate the control of the NHSE of the 1D lattice in the triangular configuration via staggered two-particle dissipation within the two-excitation subspace. This mechanism can be naturally extended to higher-excitation sectors. Here, we present the results within the three-excitation subspace.

Figure \ref{FigTrigThree}(a,e) displays the complex eigenenergies of $\hat{\mathcal{H}}_\text{nH}$ with the three-excitation subspace for different values of $\kappa$ under OBCs (blue dots) and  PBCs (red dots). The eigenspectrum consists of a continuum of scattering states and discrete triplon bands. The triplon bands correspond to bound three bosons occupying the same site. We focus on two triplon bands, highlighted on the far left of Fig.~\ref{FigTrigThree}(a,e) and shown in detail in Fig.~\ref{FigTrigThree}(b1,b2,f1,f2). In the complex plane, each triplon band, marked by a magenta star or diamond, forms a point gap under PBCs (red dots). These point gaps enclose the corresponding eigenmodes under OBCs (blue dots), indicating the presence of the NHSE.

The site-resolved particle densities of the two triplon bands are plotted in Fig.~\ref{FigTrigThree}(c,d) and Fig.~\ref{FigTrigThree}(g,h) for $\kappa/t' = 4$ and $\kappa/t' = 16$, respectively. The triplons originating from two distinct bands become localized at the boundary due to the NHSE.   Most remarkably, two-particle dissipation can reverse the skin-mode localization direction of one triplon band [see Fig.~\ref{FigTrigThree}(h)] as two-particle loss $\kappa$ increases due to the interplay of staggered two-particle loss, synthetic magnetic flux, and long-range hopping in the interacting systems.

\textit{{\color{blue}Winding number}}.---The NHSE originates from intrinsic non-Hermitian topology. In order to characterize the point-gap topology, we introduce a twist angle $\varphi$ ($\varphi \in [0,~2\pi]$) to PBC, and   define the many-body winding number \cite{PhysRevB.106.L121102,PhysRevB.105.165137,PhysRevB.106.205147}
\begin{align}\label{winding}
	\mathcal{W} = \oint_{0}^{2\pi} \frac{d \varphi}{2\pi i} \frac{\partial}{\partial \varphi} \log\left[\det \left(\hat{\mathcal{H}}_\text{nH} -E_\textrm{ref}\right)\right], 
\end{align}
where $E_\textrm{ref}$ is the complex reference energy inside the point gap. 

For two-excitation subspace, the nonzero winding number $\mathcal{W} = -2$ for each doublon band reflects the intrinsic topological origin of the doublons' NHSE for $\kappa / t' = 4$ in Fig.~\ref{FigTrig2}(a). In contrast, the winding number of the doublon band with the larger imaginary part becomes $\mathcal{W} = 2$ for $\kappa / t' = 16$ in Fig.~\ref{FigTrig2}(e), indicating a reversal of the skin-localization direction for this band. In contrast, the other doublon band has the winding number with $\mathcal{W} = -2$ for $\kappa / t' = 16$, indicating no reversal of the skin-localization direction as the two-particle dissipation increases for the doublon band with smaller imaginary part.  

For the three-excitation subspace, each triplon band has a winding number $\mathcal{W} = -3$ at $\kappa / t' = 4$ [Fig.~\ref{FigTrigThree}(a)], reflecting its intrinsic topological origin. In contrast, at $\kappa / t' = 16$ [Fig.~\ref{FigTrigThree}(e)], the winding number of the triplon band with the larger imaginary part becomes $\mathcal{W} = 3$, indicating a reversal of the skin-mode localization direction for this band. The winding number of the triplon band with the smaller imaginary part remains $\mathcal{W} = -3$, indicating no reversal of the localization direction for this band.


%

\clearpage \widetext
\begin{center}
	\section{Supplemental Material for ``Dissipation and Interaction-Controlled Non-Hermitian Skin Effects"}
\end{center}
\setcounter{equation}{0} \setcounter{figure}{0}
\setcounter{table}{0} \setcounter{page}{1} \setcounter{secnumdepth}{3} \makeatletter
\renewcommand{\theequation}{S\arabic{equation}}
\renewcommand{\thefigure}{S\arabic{figure}}
\renewcommand{\bibnumfmt}[1]{[S#1]}
\renewcommand{\citenumfont}[1]{S#1}

\makeatletter
\def\@hangfrom@section#1#2#3{\@hangfrom{#1#2#3}}
\makeatother


\maketitle

\section{Effective non-Hermitian Hamiltonian based on post-selection measurement}	

In experiments, the effective non-Hermitian Bose-Hubbard model in our work could be tested using ultracold atoms \cite{Winkler2006SM,PhysRevLett.129.070401SM, Ren2022SM, PhysRevLett.122.023601SM,PhysRevLett.130.063001SM} and superconducting quantum circuits \cite{Roushan2016SM, Naghiloo2019SM,PhysRevLett.127.140504SM}, by continuously monitoring the particle number followed by a postselection measurement~\cite{PhysRevLett.130.063001SM,Naghiloo2019SM}. 

The lattice without dissipation is described by the Hamiltonian $\hat{\mathcal{H}}$, e.g., the 1D Bose-Hubbard model with long-range hopping shown in Fig.~1 of the main text. When the lattice is subjected to staggered onsite two-particle losses of rates $\kappa \pm \delta$ on odd and even sites, the system dynamics is governed by the Lindblad master equation \cite{Scully1997SM,Breuer2007SM,arXiv:1902.00967SM}  
\begin{align}\label{mastereq}
	\frac{d \hat{\rho}}{dt} = & -i \left[\hat{\mathcal{H}}, ~\hat{\rho}\right] + \sum_{j} \left[\kappa + (-1)^{j} \delta \right] \mathcal{D}[\hat{a}_{j} \hat{a}_{j}]\hat{\rho},
\end{align}
where $\hat{\rho}$ is the system density matrix, and the Lindblad superoperator $\mathcal{D}[\hat{\mathcal{L}}]\hat{\rho} =\hat{\mathcal{L}} \hat{\rho} \hat{\mathcal{L}}^\dagger - \{\hat{\mathcal{L}}^\dagger \hat{\mathcal{L}}, ~\hat{\rho}\}/2 $ represents bosonic dissipation. 

By continuously monitoring the particle number followed by a post-selection measurement \cite{PhysRevA.94.053615SM}, the dissipative system  in Eq.~(\ref{mastereq}) is described by the effective non-Hermitian Hamiltonian     
\begin{align}\label{NH_Heff}
	\hat{\mathcal{H}}_\text{nH} =   \hat{\mathcal{H}}   - \frac{i}{2} \sum_{j=1}^{L} \left[\kappa + (-1)^{j} \delta  \right] \hat{a}^\dagger_j \hat{a}^\dagger_j \hat{a}_j \hat{a}_j,
\end{align}
where $\kappa > \delta$	is required.

\section{Absence of NHSE Reversal Under Staggered  Single-Particle Loss}

In this section, unlike the staggered two-particle loss, we demonstrate that the reversal of the non-Hermitian skin effect (NHSE)  does not occur in the presence of staggered single-particle loss, even when combined with synthetic magnetic flux and next-nearest-neighbor hopping. We consider a one-dimensional Bose-Hubbard model, subject to the staggered single-particle loss,  arranged in a triangular lattice configuration, described by the Hamiltonian 
\begin{align}\label{H012}
	\hat{\mathcal{H}}_\text{s} = & - \sum_{j=1}^{L} t\left( \hat{a}^\dagger_{j+1} \hat{a}_j + \text{H.c.}\right) - \sum_{j=1}^{L/2} \left(t^\prime e^{-i\phi} \hat{a}^\dagger_{2j+1} \hat{a}_{2j-1}  + \text{H.c.}\right) - U \sum_{j=1}^{L} \hat{a}^\dagger_j \hat{a}^\dagger_j \hat{a}_j \hat{a}_j \nonumber \\ & 
	- \frac{i}{2} \sum_{j=1}^{L} \left[\lambda + (-1)^{j} \lambda_0 \right] \hat{n}_j .
\end{align}

\begin{figure}[!t]
	\centering
	\includegraphics[width=18cm]{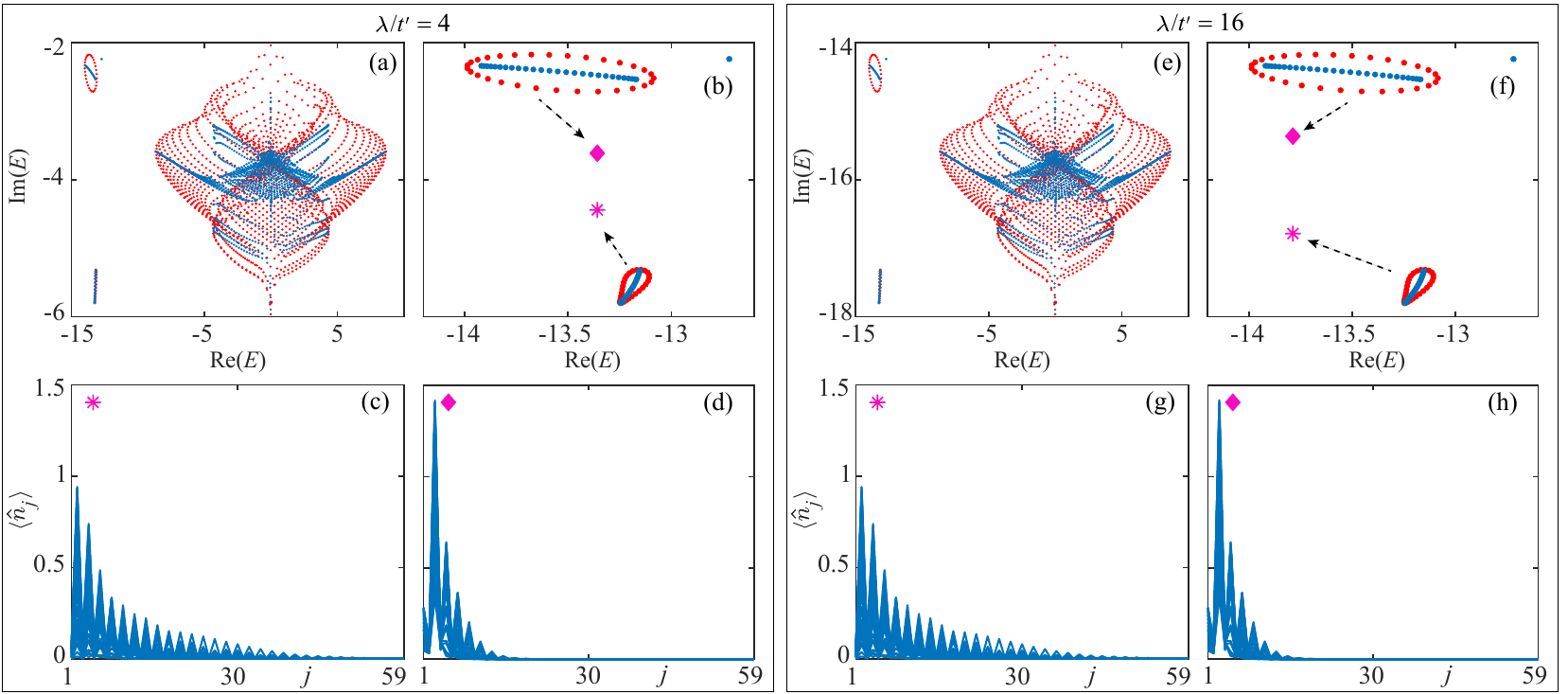}
	\caption{ Complex eigenenergies of $\hat{\mathcal{H}}_\text{s}$, in a triangular  configuration, with the staggered single-particle loss $\lambda \pm \lambda_0$ under OBCs (blue dots) and PBCs (red dots)  (a,b) for $\lambda /t' =4$ and (e,f) for $\lambda/t' =16$.  The zoom‐in view of the two doublon bands on the far left of (a,e) is shown in (b,f), where the magenta star and diamond denote the two branches of the doublon bands. The corresponding particle densities $\langle \hat{n}_j \rangle$ are shown in 	(c,) and (g,h) under OBCs.  Other parameters are: $ U/t' = 6 $,  $ t/t'=2 $,   $\lambda_0/t'=2$, $\phi = \pi/2$, and $L=59$.}\label{FigTrigS1}
\end{figure}

Figure \ref{FigTrigS1}(a,e) displays the complex eigenenergies of $\hat{\mathcal{H}}_\text{s}$ for different values of $\lambda$ under OBCs (blue dots) and  PBCs (red dots). We focus on two doublon bands, highlighted on the far left of Fig.~\ref{FigTrigS1}(a,e) and shown in detail in Fig.~\ref{FigTrigS1}(b,f). In the complex plane, each doublon band, marked by a magenta star or diamond, forms a point gap under PBCs (red dots). These point gaps enclose the corresponding eigenmodes under OBCs (blue dots), indicating the presence of the NHSE.

The site-resolved particle densities of the two doublon bands are plotted in Fig.~\ref{FigTrigS1}(c,d) and Fig.~\ref{FigTrigS1}(g,h) for $\lambda/t' = 4$ and $\lambda/t' = 16$, respectively. In both cases, the doublon modes are localized at the left boundary, further confirming the NHSE. This behavior arises from the interplay between the effective magnetic flux experienced by doublons and the staggered particle loss between odd and even sites, sharing the same mechanism as in the noninteracting case \cite{PhysRevLett.124.250402SM,PhysRevLett.129.070401SM}. However, in contrast to the case of staggered two-particle loss, the direction of localization is not reversed even under large single-particle loss in Fig.~\ref{FigTrigS1}(g,h).

\section{Effective Hamiltonian of Doublons}

We consider the strong-interaction limit with $\abs{U} \gg \abs{t}, \abs{t^\prime}$ in the double-excitation subspace. In this case, the doublons are tightly bound, where their two-particle components  reside on the same site. Moreover,  the doublon states are energetically well-separated from  two-particle scattering states.  To derive the effective Hamiltonian $\hat{\mathcal{H}}_\text{eff}$ of doublons, we split the Hamiltonian $\hat{\mathcal{H}}_\text{nH}$ into its dominant term  
\begin{align}\label{H0}
	\hat{\mathcal{H}}_0 = - U \sum_{j=1}^{L} \hat{a}^\dagger_j \hat{a}^\dagger_j \hat{a}_j \hat{a}_j - \frac{i}{2} \sum_{j=1}^{L} \left[\kappa + (-1)^{j} \delta\right] \hat{a}^\dagger_j \hat{a}^\dagger_j \hat{a}_j \hat{a}_j,
\end{align}
and a perturbation  
\begin{align}\label{V}
	\hat{V} = & - \sum_{j=1}^{L} t\left( \hat{a}^\dagger_{j+1} \hat{a}_j + \text{H.c.}\right) - \sum_{j=1}^{L/2} \left(t^\prime e^{-i\phi} \hat{a}^\dagger_{2j+1} \hat{a}_{2j-1} + \text{H.c.}\right).
\end{align}

In the strong-interaction limit, the tightly bound doublons mainly reside at the same site  in the double-excitation subspace, where the double-occupied states read
\begin{align}\label{doublon}
	\ket{d_j} \equiv \frac{1}{\sqrt{2}} \hat{a}^\dagger_j \hat{a}^\dagger_j \ket{0} = \hat{d}^\dagger_{j} \ket{0}.
\end{align}

Projecting $\hat{\mathcal{H}}_0$ onto $\ket{d_j}$   in the double-occupied states, for odd and even sites, we have 
%
%
%
\begin{align}\label{doublon2}
	E_d^{\text{odd}} = -2U-i(\kappa-\delta), ~~~~ E_d^{\text{even}} = -2U-i(\kappa+\delta).
\end{align}
In addition to the  double-occupied states, the single-occupied states are written as 
\begin{align}\label{two-particle1}
	\ket{s_{j^\prime j}^1} \equiv \hat{a}^\dagger_{2j^\prime-1} \hat{a}^\dagger_{2j-1} \ket{0} ~~ \textrm{for}~~ j^\prime \neq j, ~~~ \ket{s_{j^\prime j}^2} \equiv \hat{a}^\dagger_{2j^\prime} \hat{a}^\dagger_{2j-1} \ket{0}, ~~~\ket{s_{j^\prime j}^3} \equiv \hat{a}^\dagger_{2j^\prime} \hat{a}^\dagger_{2j} \ket{0}~~ \textrm{for}~~j^\prime \neq j.
\end{align}
Projecting $\hat{\mathcal{H}}_0$ onto these single-occupied states, we have 
\begin{align}\label{doublon3}
	E_{s1} = E_{s2} = E_{s3} = 0.
\end{align}

Based on   third-order quasidegenerate perturbation theory \cite{Bir1974SM,CCohenTannoudji1AtomSM,PhysRevA.97.013637SM}, the nonzero matrix elements of the effective Hamiltonian are given by
\begin{align}\label{peturbation}
	\bra{d} \hat{\mathcal{H}}_\text{eff}\ket{d^\prime} &= E_d \delta_{d,d^\prime} + \bra{d} \hat{V} \ket{d^\prime} + \frac{1}{2} \sum_{s} \bra{d} \hat{V} \ket{s} \bra{s} \hat{V} \ket{d^\prime} \times \left[\frac{1}{E_d - E_s} + \frac{1}{E_{d^\prime} - E_s}\right] \notag \nonumber \\
	&~~~ + \frac{1}{2} \sum_{s s^\prime} \bra{d} \hat{V} \ket{s} \bra{s} \hat{V} \ket{s^\prime} \bra{s^\prime} \hat{V} \ket{d^\prime} \times \left[\frac{1}{(E_d - E_s)(E_d - E_{s^\prime})} + \frac{1}{(E_{d^\prime} - E_s)(E_{d^\prime} - E_{s^\prime})}\right].
\end{align}

Inserting Eqs.~(\ref{H0}-\ref{doublon3}) into Eq.~(\ref{peturbation}), we achieve the effective Hamiltonian of doublon bands
\begin{align}\label{Hamileff1}
	\hat{\mathcal{H}}_\text{eff} =& - \sum_{j=1}^{L-1} \left(t_\text{R} \hat{d}^\dagger_{j+1} \hat{d}_j + t_\text{L} \hat{d}^\dagger_{j} \hat{d}_{j+1}\right) - \sum_{j=1}^{L/2-2} \left(J_\text{R} \hat{d}^\dagger_{2j+1} \hat{d}_{2j-1} + J_\text{L} \hat{d}^\dagger_{2j-1} \hat{d}_{2j+1} \right)  \nonumber \\
	&  + U_\text{odd} \sum_{j=1}^{L/2} \hat{d}^\dagger_{2j-1} \hat{d}_{2j-1} + U_\text{even} \sum_{j=1}^{L/2} \hat{d}^\dagger_{2j} \hat{d}_{2j}.
\end{align}

In Eq.~(\ref{Hamileff1}), $t_\text{L}$ and $t_\text{R}$ are the nearest-neighbor  hopping amplitudes of doublons, which are given by 
\begin{align}\label{tLBulkeffect}
	t_\text{L} =  \bra{d_{j}} \hat{\mathcal{H}}_\text{eff} \ket{d_{j+1}} 
	=  \frac{2 (2U+i\kappa) t^2}{(2U+i\kappa)^2+\delta^2} + \frac{6\left[ (2U+i\kappa)^2-\delta^2 \right] t^2 t^\prime e^{i\phi} }{\left[(2U+i\kappa)^2 + \delta^2 \right]^2} ,
\end{align}
\begin{align}\label{tRBulkeffect}
	t_\text{R} =  \bra{d_{j+1}} \hat{\mathcal{H}}_\text{eff} \ket{d_j}
	=  \frac{2 (2U+i\kappa) t^2}{(2U+i\kappa)^2+\delta^2} + \frac{6\left[ (2U+i\kappa)^2-\delta^2 \right] t^2 t^\prime e^{-i\phi} }{\left[(2U+i\kappa)^2 + \delta^2 \right]^2} .
\end{align}

The next-nearest-neighbor hopping amplitudes $J_\text{R/L}$ of doublons are derived as   
\begin{align}\label{JRoddeffect}
	J_\text{R} =&  \bra{d_{2j+1}} \hat{\mathcal{H}}_\text{eff} \ket{d_{2j-1}} = \frac{2 t^{\prime2} e^{-2i\phi}}{2U+i\kappa - i\delta} + \frac{6t^2t^\prime e^{-i\phi}}{  (2U+i\kappa- i\delta )^2},
\end{align}
\begin{align}\label{JLoddeffect}
	J_\text{L} =& \bra{d_{2j-1}} \hat{\mathcal{H}}_\text{eff} \ket{d_{2j+1}} = \frac{2 t^{\prime2} e^{2i\phi}}{2U+i\kappa - i\delta} + \frac{6t^2t^\prime e^{i\phi}}{ \left( 2U+i\kappa- i\delta \right)^2},
\end{align}

The onsite potentials $U_\text{odd}$ and $U_\text{even}$ for odd  and even sites are given by
\begin{align}\label{UoddBulkeffect}
	U_\text{odd} = \bra{d_{2j-1}} \hat{\mathcal{H}}_\text{eff} \ket{d_{2j-1}}	 =  -2U -i(\kappa-\delta) - \frac{4(t^2 + {t^\prime}^2)}{2U+i\kappa -i\delta} - \frac{8 t^2 t^\prime \cos \phi }{(2U+i\kappa - i\delta)^2},
\end{align}
\begin{align}\label{UevenBulkeffect}
	U_\text{even} = \bra{d_{2j}} \hat{\mathcal{H}}_\text{eff} \ket{d_{2j}} = -2U -i(\kappa+\delta) - \frac{4t^2}{2U+i\kappa+i\delta} - \frac{4 t^2 t^\prime \cos \phi }{(2U+i\kappa + i\delta)^2}.
\end{align}

According to Eq.~(\ref{peturbation}), under OBCs, the nearest-neighbor hopping amplitudes  and onsite potentials involving boundary sites should be modified, and they are given by
\begin{align}\label{ULoddedgeeffect}
	U_\text{edge} = \bra{d_{j}} \hat{\mathcal{H}}_\text{eff} \ket{d_{j}}  = -2U -i(\kappa-\delta) - \frac{2(t^2 + {t^\prime}^2)}{2U+i\kappa -i\delta} - \frac{4 t^2 t^\prime \cos \phi }{(2U+i\kappa - i\delta)^2}, ~~~ \textrm{for} ~~~ j =1,L.
\end{align}

\begin{figure*}[!t]
	\centering
	\includegraphics[width=18cm]{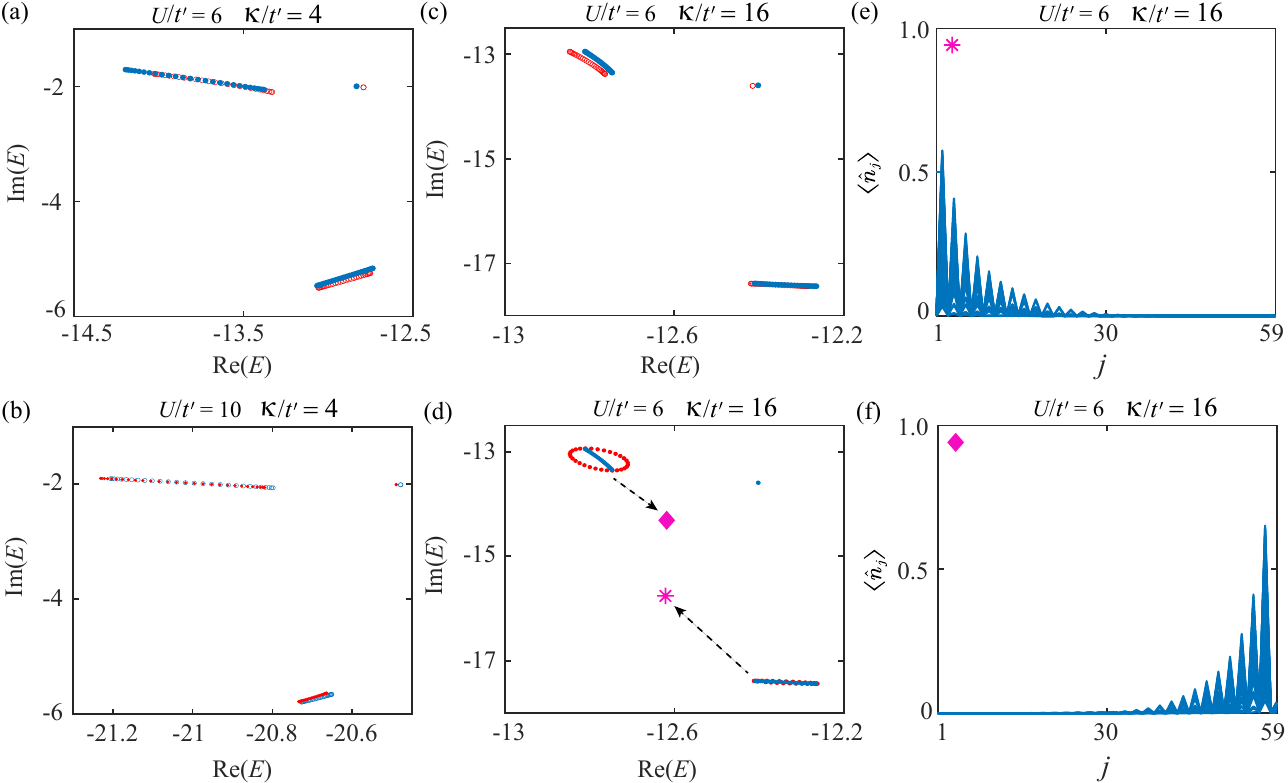}
	\caption{(a-c) Complex eigenenergies of doublons, in the triangular configuration, calculated using $\hat{\mathcal{H}}_\text{nH}$ (red circles) and the effective Hamiltonian $\hat{\mathcal{H}}_\text{eff}$ (blue dots) in the strong-interaction limit under OBCs. The doublon spectra are well captured by the effective Hamiltonian $\hat{\mathcal{H}}_\text{eff}$ for the large $U$.  (d) Doublon bands of $\hat{\mathcal{H}}_\text{eff}$ under OBCs (blue dots) and PBCs (red dots), where the magenta star and diamond denote the two branches of the doublon bands. The corresponding particle densities $\langle \hat{n}_j \rangle$ using $\hat{\mathcal{H}}_\text{eff}$ are shown in 	(e) and (f).  Other parameters are     $ t/t'=2 $,   $\delta/t'=2$, $\phi = \pi/2$, and $L=59$. }\label{FigTrigS2}
\end{figure*}

In summary, for the strong-interaction limit, we obtain the effective Hamiltonian of doublons with 
\begin{align}\label{Hamileff2}
	\hat{\mathcal{H}}_\text{eff} =& -  \sum_{j=1}^{L-1} \left( t_R \hat{d}^\dagger_{j+1} \hat{d}_j +  t_L \hat{d}^\dagger_{j} \hat{d}_{j+1}\right) - \sum_{j=1}^{L/2-2} \left(J_\text{R} \hat{d}^\dagger_{2j+1} \hat{d}_{2j-1} + J_\text{L} \hat{d}^\dagger_{2j-1} \hat{d}_{2j+1} \right)  \nonumber \\
	& + U_1 \sum_{j=1}^{L/2} \hat{d}^\dagger_{2j-1} \hat{d}_{2j-1} + U_2 \sum_{j=1}^{L/2} \hat{d}^\dagger_{2j} \hat{d}_{2j},
\end{align}
where 
\begin{align}\label{tLBulkeffectnew}
	t_L = \frac{2 (2U+i\kappa) t^2}{(2U+i\kappa)^2+\delta^2} + \frac{6\left[ (2U+i\kappa)^2-\delta^2 \right] t^2 t^\prime e^{i\phi} }{\left[(2U+i\kappa)^2 + \delta^2 \right]^2},
\end{align}
\begin{align}\label{tRBulkeffectnew}
	t_R = \frac{2 (2U+i\kappa) t^2}{(2U+i\kappa)^2+\delta^2} + \frac{6\left[ (2U+i\kappa)^2-\delta^2 \right] t^2 t^\prime e^{-i\phi} }{\left[(2U+i\kappa)^2 + \delta^2 \right]^2},
\end{align}
\begin{align}\label{JRoddeffectspy}
	J_\text{R} = \frac{6t^2t^\prime }{  (2U+i\kappa- i\delta )^2} e^{-i\phi} +  \frac{2 t^{\prime2} }{2U+i\kappa - i\delta} e^{-2i\phi},
\end{align}
\begin{align}\label{JLoddeffectspy}
	J_\text{L} =   \frac{6t^2t^\prime }{  (2U+i\kappa- i\delta )^2} e^{i\phi} +  \frac{2 t^{\prime2} }{2U+i\kappa - i\delta} e^{2i\phi},
\end{align}
\begin{align}\label{UoddBulkeffect2}
	U_1   = -2U -i(\kappa-\delta) - \frac{4(t^2 + {t^\prime}^2)}{2U+i\kappa -i\delta} - \frac{8 t^2 t^\prime \cos \phi }{(2U+i\kappa - i\delta)^2},
\end{align}
\begin{align}\label{UevenBulkeffect2}
	U_2 = -2U -i(\kappa+\delta) - \frac{4t^2 }{2U+i\kappa+i\delta} - \frac{4 t^2 t^\prime \cos \phi }{(2U+i\kappa + i\delta)^2},
\end{align}
where $U_1$ and $U_2$ are the onsite potential, excluding contributions from the two boundary sites.

We plot the complex eigenenergies of doublon bands calculated using $\hat{\mathcal{H}}_\text{nH}$ (red circles) and the effective Hamiltonian $\hat{\mathcal{H}}_\text{eff}$ (blue dots) in the strong-interaction limit under OBCs, as shown in Fig.~\ref{FigTrigS2}(a,b,c). The effective Hamiltonian $\hat{\mathcal{H}}_\text{eff}$ exhibits a well approximation to the exact results from $\hat{\mathcal{H}}_\text{nH}$ for the large $U$. The site-resolved particle density $\langle \hat{n}_j \rangle$ of doublons, corresponding to the zoom-in spectrum in  Fig.~\ref{FigTrigS2}(d), is shown in Fig.~\ref{FigTrigS2}(e,f), where state distributions are well captured by the effective Hamiltonian.

\begin{figure}[!tb]
	\centering
	\includegraphics[width=17.5cm]{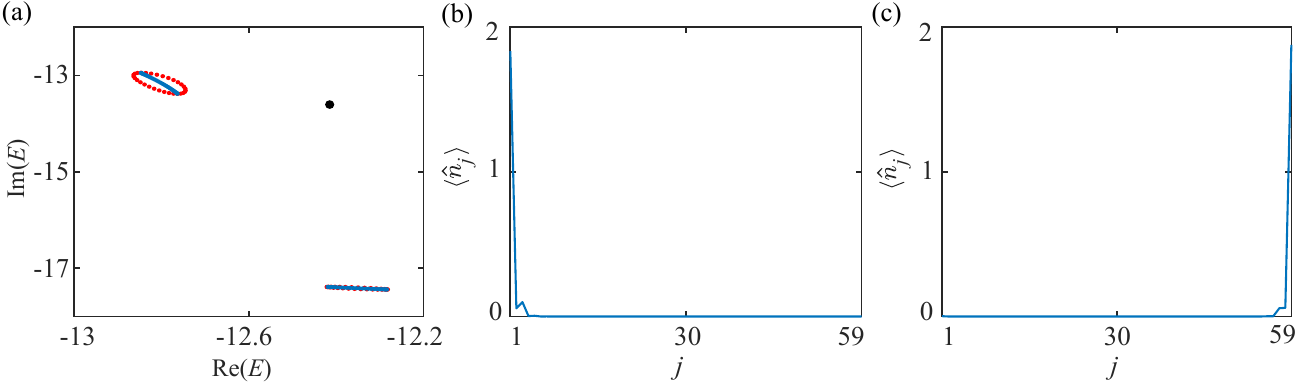}
	\caption{(a) Complex eigenenergies of doublons under OBCs (blue dots) and PBCs (red dots), where the black dot denotes two degenerate Tamm-Shockley states under OBCs. The corresponding particle densities $\langle \hat{n}_j \rangle$ are shown in 	(b) and (c) under OBCs.  Other parameters are: $ U/t' = 6 $,  $\kappa/t'=16$, $ t/t'=2 $,      $\delta/t'=2$, $\phi = \pi/2$, and $L=59$.}\label{FigTrigS4}
\end{figure}

With open boundary conditions, the renomalized onsite potentials in the boundary and bulk sites are different. Inside the bulk sites, the onsite potentials are given in Eqs.~(\ref{UoddBulkeffect2}) and (\ref{UevenBulkeffect2}). At the boundary sites, the onsite potentials are derived from Eq.~(\ref{ULoddedgeeffect})    as
\begin{align}\label{ULoddedgeeffect2}
	U_\text{edge} =& -\left(2U + \frac{4U( t^2 + {t^\prime}^2)}{4U^2+(\kappa-\delta)^2} + \frac{4t^2 t^\prime (4 U^2 -(\kappa-\delta)^2)\cos(\phi)}{(4U^2+(\kappa- \delta)^2)^2}\right) \nonumber \\
	& + i \left[\frac{2(\kappa-\delta)(t^2 + {t^\prime}^2)}{4U^2+(\kappa-\delta)^2}  + \frac{16t^2 t^\prime U(\kappa-\delta)\cos(\phi)}{(4U^2+(\kappa-\delta)^2)^2} - (\kappa - \delta)\right], ~~\text{for} ~~ j =1,L.
\end{align}
Due to the different onsite potentials between boundary and bulk sites (magenta diamond and star),  two isolated states of doublons, lying outside the point gaps, are localized at either the left or right boundaries [see  Fig.~\ref{FigTrigS4}]. There are called two-particle Tamm-Shockley states \cite{Shockley1932SM,PhysRev.56.317SM}, which have no topological origin.

\begin{figure}[!b]
	\centering
	\includegraphics[width=12cm]{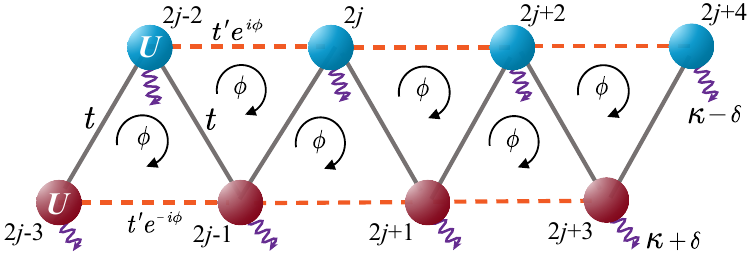}
	\caption{Schematic showing a 1D Bose-Hubbard model with long-range hopping, arranged in a two-leg zigzag configuration. The onsite interaction is denoted by $U$, while  staggered two-particle losses between two legs, with rates $\kappa \pm \delta$, are indicated by purple wavy lines. The parameters $t$ and $t^{\prime}$ represent the nearest-neighbor (dark solid lines) and next-nearest-neighbor (dashed orange lines) hopping amplitudes, respectively. The complex next-nearest-neighbor hopping introduces a synthetic magnetic flux $\phi$ that threads each triangular plaquette. }\label{FigZigZag1}
\end{figure}

\section{1D  lattice in zigzag configuration}

In this section, we consider   a 1D Bose-Hubbard model with long-range hopping, arranged in a two-leg zigzag configuration. As shown in Fig.~\ref{FigZigZag1}, the 1D zigzag lattice is  featured by long-range next-nearest-neighbor hopping, onsite many-body interactions, and staggered two-particle dissipation. The phase associated with the long-range hopping introduces a synthetic magnetic flux $\phi$, which threads through each triangular plaquette.   The effective non-Hermitian Hamiltonian of the system is written as
\begin{align}\label{NH_HeffZigzag}
	\hat{\mathcal{H}}_\text{nH} = & -\sum_{j=1}^{L/2-1} \left(t^\prime e^{i\phi} \hat{a}^\dagger_{2j+2} \hat{a}_{2j} + t^\prime e^{-i\phi} \hat{a}^\dagger_{2j+1} \hat{a}_{2j-1} + \text{H.c.}\right)   - \sum_{j=1}^{L-1} t\left( \hat{a}^\dagger_{j+1} \hat{a}_j + \text{H.c.}\right) - U  \sum_{j=1}^L \hat{a}^\dagger_j \hat{a}^\dagger_j \hat{a}_j \hat{a}_j     \nonumber \\ & - \frac{i}{2} \sum_{j=1}^{L} \left[\kappa - (-1)^{j} \delta  \right] \hat{a}^\dagger_j \hat{a}^\dagger_j \hat{a}_j \hat{a}_j.
\end{align}

\begin{figure*}[!tb]
	\centering
	\includegraphics[width=18cm]{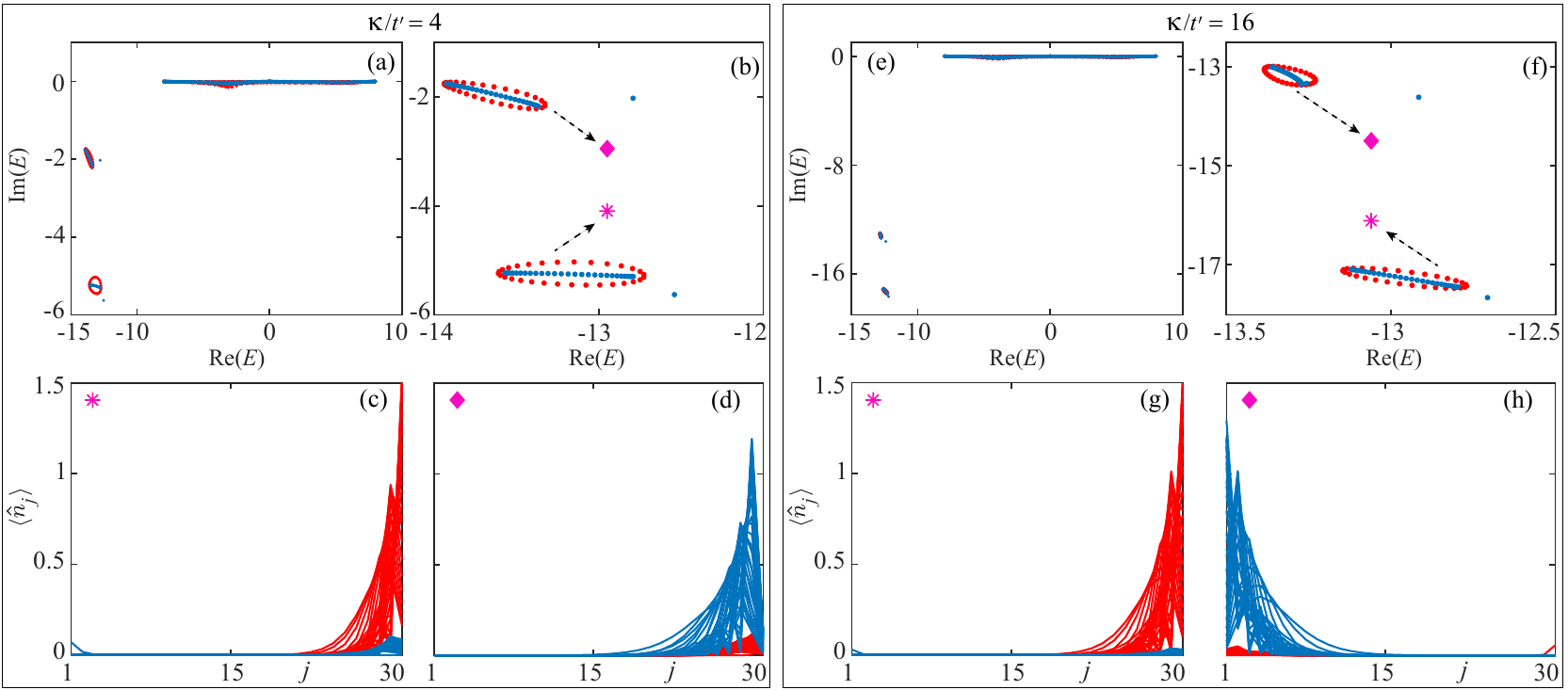}
	\caption{ Complex eigenenergies of $\hat{\mathcal{H}}_\text{nH}$, in the zigzag configuration, under OBCs (blue dots) and PBCs (red dots) (a) for $\kappa/t'=4$ and  (e) for $\kappa/t'=16$.  The zoom‐in view of the two doublon bands on the far left of (a,e) is shown in (b,f), where the magenta star and diamond denote the two branches of the doublon bands. The corresponding particle densities $\langle \hat{n}_j \rangle$ are shown in 	(c,d) and (g,h) under OBCs, where the  red (blue) curves indicate the state  distributions at the odd (even) sites.  Other parameters are: $ U/t' = 6 $,  $ t/t'=2 $,  $\delta/t'=2$, $\phi = \pi/2$, and $L=60$.}\label{FigZigZag2}
\end{figure*}

We plot the complex  eigenenergies of $\hat{\mathcal{H}}_\text{nH}$, in the zigzag configuration, for different      $\kappa$ under  OBCs (blue dots) and PBCs (red dots) in Fig.~\ref{FigZigZag2}(a,e). The eigenspectrum consists of a continuum of scattering states and discrete doublon bands. The scattering states are superpositions of two particles on different sites, while the doublon bands correspond to bound pairs occupying the same site, with eigenenergies well separated from the scattering continuum [see Fig.~\ref{FigZigZag2}(a,e)]. We focus on two such doublon bands, highlighted on the far left of Fig.~\ref{FigZigZag2}(a,e) and magnified in Fig.~\ref{FigZigZag2}(b,f). In the complex plane, each doublon band, marked by a magenta star or diamond, forms a point gap (red dots) under PBCs.

Figures \ref{FigZigZag2}(c,d) and \ref{FigZigZag2}(g,h) show the site-resolved particle densities of the two doublon bands for $\kappa/t' = 4$ and $\kappa/t' = 16$, respectively, with red (blue) curves indicating the occupation at odd (even) sites.  All the doublons are localized at the boundary, indicating the NHSE. Moreover, the states from two branches of doublon bands occupy different legs of the zigzag geometry.   Most remarkably, two-particle dissipation can reverse the skin-mode localization direction of one doublon band [see Fig.~\ref{FigZigZag2}(d,h)]. Previous studies have shown that staggered single-particle loss, combined with magnetic flux in the zigzag configuration, can induce the NHSE by favoring transport along lower-loss sites, but it cannot reverse the effect.  Such reversal does not occur in interacting systems with staggered single-particle loss.  	These results reveal that staggered two-particle loss constitutes a fundamentally novel mechanism for inducing and reversing the NHSE, distinct from previously reported scenarios, such as the multiple-hopping-pathway interference mechanism proposed for NHSE reversal.

%
%

In order to  understand the underlying hidden mechanism of the unexpected reversal of the  localization direction of the doublons in the zigzag configuration, we derive the effective Hamiltonian $\hat{\mathcal{H}}_\text{eff}$ describing the doublon bands in the strong-interaction limit with   $\abs{U} \gg \abs{t}, \abs{t^\prime}$.   Based on  third-order quasidegenerate  perturbation theory, we obtain the effective Hamiltonian $\hat{\mathcal{H}}_\text{eff}$ of doublons as
\begin{align}\label{Hamileff20}
	\hat{\mathcal{H}}_\text{eff} = 	& - t_0 \sum_{j} \left( \hat{d}^\dagger_{j+1} \hat{d}_j +   \textrm{H.c.}\right) + \sum_{j} \left(U_1 \hat{n}_{2j-1} + U_2 \hat{n}_{2j}  \right)    \nonumber \\ 	
	& - \sum_{j} \left(J_\text{1,R} \hat{d}^\dagger_{2j+1} \hat{d}_{2j-1} + J_\text{1,L} \hat{d}^\dagger_{2j-1} \hat{d}_{2j+1} \right)  \nonumber \\
	& - \sum_{j} \left(J_\text{2,R} \hat{d}^\dagger_{2j+2} \hat{d}_{2j} + J_\text{2,L} \hat{d}^\dagger_{2j} \hat{d}_{2j+2} \right),
\end{align}
where $J_\text{1,R}=J(\delta, \phi)$, $J_\text{1,L}=J(\delta, -\phi)$, $J_\text{2,R}=J(-\delta, -\phi)$, $J_\text{2,L}=J(-\delta, \phi)$, $U_1 = 	U(\delta)$,   $U_2 = U(-\delta)$, and  
\begin{align}\label{tLBulkeffectnew0}
	t_0 = \frac{2 (2U+i\kappa) t^2}{(2U+i\kappa)^2+\delta^2} + \frac{12\left[ (2U+i\kappa)^2-\delta^2 \right] t^2 t^\prime \cos \phi }{\left[(2U+i\kappa)^2 + \delta^2 \right]^2}  ,
\end{align}
\begin{align}\label{JRoddeffectspy02}
	J(\delta, \phi) = \frac{6t^2 t^\prime}{(2U+i\kappa+i\delta)^2}  e^{-i\phi} + \frac{2 t^{\prime2}}{2U + i\kappa + i\delta} e^{-2i\phi},
\end{align}
\begin{align}\label{UoddBulkeffect20}
	U(\delta)   = -2U -i(\kappa+\delta) - \frac{4(t^2 + {t^\prime}^2) }{2U+i\kappa+i\delta} - \frac{12 t^2 t^\prime \cos \phi }{(2U+i\kappa + i\delta)^2}.
\end{align}

\begin{figure}[!tb]
	\centering
	\includegraphics[width=17.8cm]{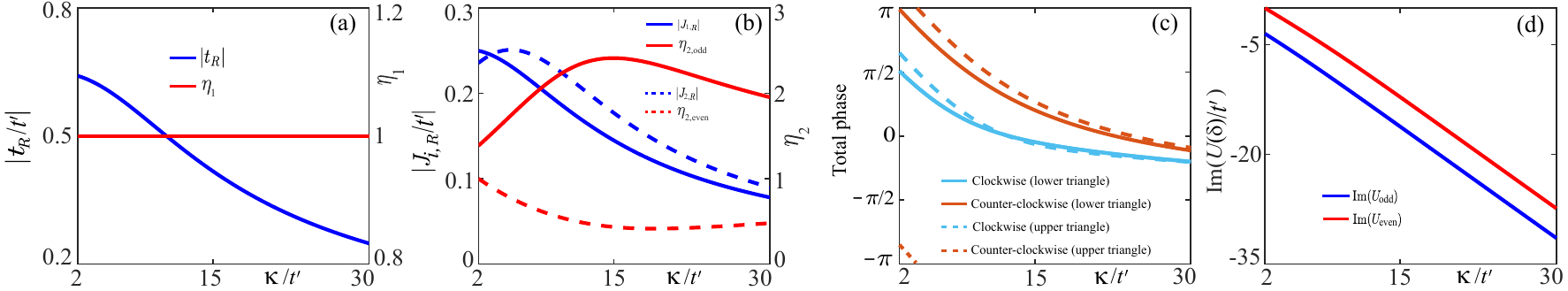}
	\caption{ (a) $\abs{t_R/t'}$ and $\eta_1 = \abs{t_R/t_L}$ versus $\kappa/t'$, (b) $\abs{J_{1,R}/t'}$ and $\eta_{2, \textrm{odd}} = \abs{J_{1,R}/J_{1,L}}$ as well as $\abs{J_{2,R}/t'}$ and $\eta_{2, \textrm{even}} = \abs{J_{2,R}/J_{2,L}}$ versus $\kappa/t'$, (c)   total phase threading each triangular plaquette versus $\kappa/t'$, and (d) $U(\delta)/t'$   versus $\kappa/t'$. Other parameters are  $ U/t' = 6 $,  $ t/t'=2 $,  $\phi = \pi/2$, and $\delta/t'=2$.}\label{FigZigZag_effectHoppingStrength}
\end{figure}

According to Eq.~(\ref{JRoddeffectspy02}), the virtual second-order hopping process of particle pairs contributes an effective phase of $2\phi$, while the third-order process contributes a phase of $\phi$. Therefore, the synthetic magnetic flux and long-range hopping, combined with onsite many-body interactions, can lead to strong nonreciprocal hopping of doublons, in the zigzag configuration, with $\abs{J_{i,\text{R}} }\neq \abs{J_{i,\text{L}}}$ ($i=1,2$) in Eq.~(\ref{JRoddeffectspy02}).

Figure \ref{FigZigZag_effectHoppingStrength} shows $\abs{t_R}$ and its ratio $\eta_1 = \abs{t_R / t_L}$, $\abs{J_{1,R}}$ and its ratio $\eta_{2, \textrm{odd}} = \abs{J_{1,R}/J_{1,L}}$,  $\abs{J_{2,R}}$ and its ratio $\eta_{2, \textrm{even}} = \abs{J_{2,R}/J_{2,L}}$, the effective magnetic flux threading the triangular lattice of doublons, and the effective staggered onsite loss as functions of the two-particle loss $\kappa$. Most remarkably, the  hopping of doublons along the leg becomes strongly nonreciprocal as $\kappa$ increases [see Fig.~\ref{FigZigZag_effectHoppingStrength}(b)]. For small $\kappa$, the   hoppings along two legs are only slightly nonreciprocal. In this regime, the right-edge skin modes are caused primarily by the effective magnetic flux and the staggered onsite loss at even and odd sites. However, for large $\kappa$, the  hopping along the even sites becomes highly nonreciprocal, with a larger left  hopping strength $J_{2,L}$. Therefore, the doublon with the larger imaginary part (which mainly occupies the even sites, as shown in Fig.~\ref{FigZigZag2}) becomes reversely localized toward the left edge.


%

\end{document}